\documentclass[dvips]{arxstspdf}
\usepackage{flushend}
\usepackage{stfloats}

\volume{24}
\issue{3}
\pubyear{2009}
\firstpage{273}
\lastpage{279}
\doi{10.1214/09-STS277B}
\referstodoi{10.1214/09-STS277}



\begin{document}
\begin{frontmatter}

\vspace*{6pt}

\title{Discussion of Likelihood Inference for Models with Unobservables:
Another View}
\runtitle{Discussion}
\pdftitle{Discussion on Likelihood Inference for Models with
Unobservables: Another View by Y. Lee and J. A. Nelder}

\begin{aug}
\author[a]{\fnms{Geert} \snm{Molenberghs}\corref{}\ead[label=e1]{geert.molenberghs@uhasselt.be}},
\author[b]{\fnms{Michael G.} \snm{Kenward}}
\and
\author[c]{\fnms{Geert} \snm{Verbeke}}
\runauthor{G. Molenberghs, M. G. Kenward and G. Verbeke}

\affiliation{Universiteit Hasselt,
Katholieke Universiteit Leuven and
London School of Hygiene and Tropical Medicine}

\address[a]{Geert Molenberghs is Professor of Biostatistics, I-BioStat, Universiteit
Hasselt, B-3590 Diepenbeek, Belgium
and I-BioStat, Katholieke Universiteit Leuven, B-3000 Leuven, Belgium.}

\address[b]{Michael G. Kenward is Professor of Medical Statistics,
Medical Statistics Unit, London School of Hygiene and Tropical
Medicine, London WC1E7HT, UK.}

\address[c]{Geert Verbeke is Professor of Biostatistics, I-BioStat, Katholieke Universiteit
Leuven, B-3000 Leuven, Belgium and
I-BioStat, Universiteit Hasselt, B-3590 Diepenbeek, Belgium.}

\end{aug}



\end{frontmatter}

\section{Introduction}

We are grateful for the opportunity to comment on Professors Lee and
Nelder's work. Like their other, related papers, they are replete with
important ideas and thought-provoking theses. We will take up just a
few of the topics touched upon and offer some reflections. In
Section~\ref{unobservables}, we consider models with, and inferences
for, unobservables. Section~\ref{cauchy} revisits one of the
counterexamples discussed by the authors which turns out to have
insightful connections with the Cauchy distribution. Connections
between generalized estimating equations and fully specified joint
distributions are touched upon in Section~\ref{gee}. Finally, the
position of the authors' computational proposals among alternative
routes is explored in Section~\ref{computation}.

\section{The Nature of Unobservables}\label{unobservables}

Although Lee and Nelder claim, in their Section~4.4, that unobservables
are often verifiable because the unobservables are latent variables for
observed data, we would like to issue a word of caution. Evidently,
using the authors' notation, variance components $\lambda$ and $\phi$
are identifiable from the data, whenever there is replication within
the $i$ levels (e.g., repeated measures $j$ on subject $i$, or litter
mates $j$ corresponding to dam $i$). However, such data-based
verification is confined, in the strict sense, to the way the {\em
induced marginal model} is accurate. This model is of a
compound-symmetry type:
%
\begin{equation}
\label{margmodln}
\mathbf{Y}_i\sim N(\mathbf{1}\beta,V_i=\lambda J_{n_i} +
\phi I_{n_i}).
\end{equation}
In other words, one can assess from the data whether $V_i$ is an
accurate description of the variance--covariance structure, just as one
can verify whether the implied constant mean is adequate. But whether
this provides an accurate description of the unobservables is another
matter because there is a many-to-one map of hierarchical models to
(\ref{margmodln}).

\begin{table*}[b]
\tabcolsep=0pt
\caption{}\label{tab1}
\begin{tabular*}{300pt}{@{\extracolsep{4in minus 4in}}lcccccc@{}}
\hline
&&$\sigma^2$&&$d$&&$2\tau$\\
\hline
$\operatorname{var}(Y_{ij})$ & $=$ & $\nu^2$& $+$ &$(\lambda^2+2\nu^2+2\nu\alpha\sqrt
{\lambda^2+\nu^2})$ &$+$ &$(-2\nu^2-2\nu\alpha\sqrt{\lambda^2+\nu
^2})$
\\
$\operatorname{cov}(Y_{ij},Y_{ik})$&$=$&&&$(\lambda^2+2\nu^2+2\nu\alpha\sqrt
{\lambda^2+\nu^2})$&$+$&$(-2\nu^2-2\nu\alpha\sqrt{\lambda^2+\nu^2})$\\
\hline
\end{tabular*}
\end{table*}

To see this more clearly, we start from this com\-pound-symmetry setting
and take an example of Mo\-lenberghs and Verbeke (\citeyear{2009Molenberghs}). Consider the
following random-intercepts model:
%
\begin{equation}\label{rimodel}
Y_{ij}=\mathbf{x}_{ij}'\bolds{\xi}+b_i+\varepsilon_{ij},
\end{equation}
where $Y_{ij}$ is the response for member $j=1,\dots,n_i$ of cluster
$i=1,\dots,N$, $\mathbf{x}_{ij}$ is a vector of known covariate values,
$\bolds{\xi}$ is a vector of unknown regression coefficients and
$b_i\sim
N(0,\lambda^2)$ is a cluster-specific random effect assumed to be
independently distributed from the residual error components
$\varepsilon_{ij}\sim N(0,\nu^2)$. The implied marginal model is
obtained by integrating (\ref{rimodel}) over the random effects.
Grouping the $Y_{ij}$ into a vector $\mathbf{Y}_i$ and assembling the
rows $\mathbf{x}
_{ij}'$ into a matrix $X_i$, this marginal distribution is
%
\begin{equation}
\label{margmod}
\mathbf{Y}_i\sim N(X_i\bolds{\xi},\lambda^2 J_{n_i}+\nu^2 I_{n_i}),
\end{equation}
in which $I_{n_i}$ denotes the identity matrix of dimension $n_i$, and
where $J_{n_i}$ equals the $n_i \times n_i$ matrix containing only
ones. Evidently, (\ref{margmodln}) is a particular instance of (\ref
{margmod}).

Traditionally, there have been two views regarding the variance
component $\lambda^2$ in the above model. In the first, where the
focus is entirely on the resulting marginal model (\ref{margmod}),
negative values for $\lambda^2$ are perfectly acceptable (Nelder,
\citeyear{1954Nelder}; Verbeke and Molenberghs, \citeyear{2000Verbeke}, Section~5.6.2) because this
merely corresponds to the occurrence of negative within-cluster
correlation $\rho=\lambda^2/(\lambda^2+\nu^2)$. In such a case, the
only requirement is that $\lambda^2 + \nu^2 > 0$ and $V_i=\lambda^2
J_{n_i}+\nu^2 I_{n_i}$ is positive definite. In the second view, when
the link between the marginal model (\ref{margmod}) and its generating
hierarchical model (\ref{rimodel}) is preserved, thereby including the
concept of random effects $b_i$ and perhaps even requiring inferences
for them, it has been considered imperative to restrict $\lambda^2$ to
nonnegative values.

In such a view, it is implicit that any hierarchical model,
corresponding to the compound-symmetry model (\ref{margmod}), should
be of the form (\ref{rimodel}). But this is not the case. To see this,
we first reiterate that the hierarchical model, corresponding to a
given marginal model, is nonunique. This originates from the random
effects' latency and is crucial to the theme of Lee and Nelder's
current paper.

To illustrate this nonuniqueness, consider the simple but insightful
case of two measurements per subject, that is, $n_i=2$. We contrast two
models, the first one of the form (\ref{rimodel}) with random
intercepts $b_i\sim N(0,\lambda^2)$ and heterogeneous errors
$\varepsilon_{ij}\sim N(0,\nu^2_j)$, $(j=1,2)$. The second takes the
following form:
%
\begin{equation}\label{secondmodel}
Y_{ij}=\mathbf{x}_{ij}'\bolds{\xi}+b_{0i}+b_{1i}(j-1)+\varepsilon_{ij}
\end{equation}
with two uncorrelated random effects,
\[
\pmatrix{
b_{0i}\cr
b_{1i}
}
\sim
N\left[
\pmatrix{
0\cr 0
},
\pmatrix{
\lambda_1^2& 0 \cr
0 & \lambda_2^2
}
\right]
\]
and homogeneous error $\varepsilon_{ij}\sim N(0,\nu^2)$.
The marginal means are, obviously, equal. At the same time, the
marginal variance--covariance matrix in the first\break model is
\begin{eqnarray}
V^{(1)} &=&
\pmatrix{
1\cr
1
}
 (\lambda^2)  \pmatrix{ 1 & 1} +
 \pmatrix{ \nu_1^2&0\cr
 0&\nu_2^2\nonumber
}
\\[-8pt]\\[-8pt]
&=&
\pmatrix{\lambda^2+\nu_1^2&\lambda^2\cr
 \lambda^2&\lambda
^2+\nu_2^2
},\nonumber
\end{eqnarray}
and the counterpart for the second model is
%
\begin{eqnarray}
 V^{(2)} &=&
\pmatrix{
1&0\cr
1&1
}
\pmatrix{
\lambda^2_1&0\cr
0&\lambda^2_2
}
\pmatrix{
1&1\cr
0&1
}\nonumber
\\
&&{}+ \pmatrix{\nu^2&0\cr
0&\nu^2
}
 \\
 &=&
\pmatrix{\lambda^2_1+\nu^2&\lambda^2_1\cr
\lambda^2_1&\lambda^2_1+\lambda^2_2+\nu^2
}.\nonumber
\end{eqnarray}
Evidently, $V^{(1)}$ and $V^{(2)}$ are equivalent, through the linear
relationships $\lambda_1^2=\lambda^2$, $\lambda_2^2=\nu_2^2-\nu
_1^2$, and $\nu^2=\nu_1^2$. What this means, in this case, is that
the observed heterogeneity in variance can be ascribed to either
heterogeneous residual errors or to the presence of a random slope. The
fitted marginal model, and hence the data, cannot be used to
distinguish between these two scenarios. Thus, one view is that fitting
a marginal model comes with an entire equivalence class of hierarchical
models that reduce to the given marginal model.

We now show that another simple extension of the distributional
assumptions of (\ref{rimodel}) allows for negative variance
components, while maintaining the model's random-intercepts interpretation.

We retain the random-intercepts model (\ref{rimodel}), but now with
the assumption,
%
\begin{equation}\label{additionalassumption}
\quad\pmatrix{
b_i\cr
\varepsilon_{i1}\cr
\vdots\cr
\varepsilon_{in_i}
}
\sim
N
\left[
\pmatrix{
0\cr
0\cr
\vdots\cr
0
}
,
\pmatrix{
d&\tau&\ldots&\tau\cr
\tau&\sigma^2&\ldots&0\cr
\vdots&\vdots&\ddots&\vdots\cr
\tau&0&\ldots&\sigma^2
}
\right].
\end{equation}
The induced conditional distribution of the measurement error vector,
given the random intercept, is
%
\begin{equation}\label{conditionalerror}
\bolds{\varepsilon}_i|b_i\sim N\biggl[
\frac{\tau b_i}{d}\mathbf{j}_{n_i},\frac{1}{d}(d\sigma
^2I_{n_i}-\tau^2 J_{n_i})
\biggr].
\end{equation}
Here, $\mathbf{j}_{n_i}$ is a $n_i$-vector of ones. Note that (\ref
{additionalassumption}) allows for marginally uncorrelated measurement
errors that become correlated, {\em conditional upon the random effect}.

The corresponding marginal model is
%
\begin{equation}
\label{margmod2}
\mathbf{Y}_i\sim N[X_i\bolds{\xi},(d+2\tau) J_{n_i}+\sigma^2
I_{n_i}].
\end{equation}
Starting from a conventional compound-symmetry model, it is clear that
$\sigma^2=\nu^2$ and $d+2\tau=\lambda^2$. Evidently, $d$ and $\tau
$ are not jointly identified, pointing to a collection of hierarchical
models that all yield the same marginal model and hence are
indistinguishable based on the data alone. To define the range of this
collection, it is necessary for
$\tau^2\le d\sigma^2$. Together with $\tau=(\lambda^2-d)/2$, this
leads to the set of allowable solutions,
%
\begin{equation}\label{drange}
d=\lambda^2+2\nu^2+2\nu\alpha\sqrt{\lambda^2+\nu^2},
\end{equation}
with $\alpha\in[-1,1]$.
From (\ref{drange}), we find
%
\begin{equation}
\tau=-\bigl(\nu^2+\nu\alpha\sqrt{\lambda^2+\nu^2}\bigr).
\end{equation}
In other words, we have a decomposition of the variance and the
 covariance, as displayed in Table~\ref{tab1}.
Observe that, when $\lambda^2$ is positive, $\tau=0$ is among the
solutions; that is, this recovers the conventional random-intercepts
model with uncorrelated errors. However, when $\lambda^2<0$, then all
values for $\tau$ are necessarily negative.

This construction reduces the conventional random-intercepts model
(\ref{rimodel}) to a special case of the extended family of
hierarchical models with correlation between random effects and
measurement errors. Once again, it is clear that one can make
inferences about the random effects, {\em given\/} that one is prepared
to make strong, and unverifiable assumptions about the hierarchical
model, stemming from the many-to-one map from hierarchical models to
the implied marginal models.
In this sense, the above derivations underscore, not just that every
compound-symmetry model can be induced by a hierarchical model, but
that an entire collection of random-intercepts models fulfills this
role. These differ from each other by the degree to which the random
intercepts and measurement errors are correlated.

The above considerations focus on random effects. This is but one
example of unobservables. Like Lee and Nelder, Verbeke and Molenberghs
(\citeyear{2009Molenberghs}) argue that so-called augmented data, in the sense of
supplementing the observed data with latent, unobserved structures, is
common throughout statistics. Examples \mbox{include} models for incompletely
observed data, describing observed and unobserved outcomes alike,
random-effects models, latent class models, latent variable models,
censored survival data, etc. Heitjan and Rubin (\citeyear{1991Heitjan}) and Zhang and
Heitjan (\citeyear{2007Zhang}) have unified some of these settings in a concept called
{\em coarsening\/}, broadly referring to the fact that the observed
data are coarser than the hypothetically conceived data structures
while models target the latter. Generally, models for augmented
structures are identifiable only by virtue of making sometimes strong
but always partially unverifiable modeling assumptions. These settings
taken together are termed {\em enriched data\/} by Verbeke and
Molenberghs (\citeyear{2009Verbeke}). Of course, there is a subtle distinction between
both concepts. In the coarse-data setting, it is understood that a part
of the data would ideally be observed but are not in practice (e.g.,
actual survival time after censoring, outcomes after dropout, etc.).
Augmented data refers rather to the addition of useful but artificial
constructs to the data setting, such as random effects, latent classes
and latent variables which are never directly observed. Such
augmentations permit simple model development and represent a very
powerful tool to succinctly accommodate posited, potentially very
complex, often causal, real-world structures, a fact of which Lee and
Nelder also make use.

Verbeke and Molenberghs (\citeyear{2009Verbeke}) show that every model for enriched-data
settings can be factored as a product of two components: the first one,
termed the marginal model, is fully identifiable from the observed
data; the second one, the conditional distribution of the enriched data
given the observed data, is entirely arbitrary.
The evident consequence is that the identification of such a part can
come from assumptions only and points at the same time to the
considerable risk for conclusions to be sensitive to such assumptions,
and ultimately to the need for conducting sensitivity analyses. It
implies that such non-identified parts can be replaced arbitrarily,
without altering the fit to the observed data but with potentially huge
implications for inferences and substantive conclusions. Put simply,
one's inferential conclusions may strongly depend on such unverifiable
portions of the model.

In the missing data case, studied in more detail by Molenberghs et al.
(\citeyear{2008Molenberghs}), one could identify the second factor by requiring, for example,
that it is of the MAR type. This means that every model assuming that
the missing data are missing not at random corresponds to another
model, producing exactly the same fit to the observed data, but now
assuming that the missing data are missing at random. In the context of
a conventional linear mixed model, Verbeke and Molenberghs (\citeyear{2009Verbeke})
illustrate the implications of the result by replacing the conditional
distribution of the random effects given the data, that is, the random
effects' posterior, by two families of exponential distributions,
\mbox{special} cases of the gamma family for the sake of illustration. This
nicely supplements the above compound-symmetry model with correlated
random effects and measurement errors.

These results imply that one should not simply adopt a hierarchical
model, only because it is convenient, is in common use, etc. Rather,
one should carefully reflect on that part of the model that cannot be
critiqued by the data. One should strive for (\ref{margmodln}) better understanding
of the dependence of one's inferences on nonverifiable model
components and (\ref{rimodel}) developing sensitivity analysis tools regarding
substantive conclusions with respect to data enrichment. Generally
speaking, inferences relative to observed data only, such as
fixed-effects and variance-component parameter estimates, are
unaffected by the choice of enrichment model. However, such aspects as
empirical Bayes predictions in linear mixed models, or predictive
distributions of unobserved measurements given observed ones, strongly
rest on unverifiable modeling assumptions. This points to the need for
sensitivity analysis. Rather than fitting a single model and putting
blind belief in it, it is more reasonable to consider a discrete or
continuous set of alternative model formulations and assess how key
inferences are vulnerable to choices made. Molenberghs and Kenward
(\citeyear{2007Molenberghs}) discuss avenues for sensitivity analysis.

As soon as one is aware of the lack of identification, there is
reasonable latitude in making pragmatic identification choices. For
example, with random effects or other latent structures, one could
express a preference for conjugate priors (Lee and Nelder, \citeyear{1996Lee,2001Lee,2003Lee})
because, in the absence of identification, the convenience and
appeal of conjugacy may be invoked.

\section{Bayarri's Example and a Cauchy-Type Family of
Distributions}\label{cauchy}

Lee and Nelder, in their Section~4.2, revisit Bayarri's example. There
is something rather peculiar about it, because it is of a Cauchy type.
We will show this in what follows. Owing to the model's absence of
finite moments, it seems natural that an estimation method ought to
encounter problems. Indeed, any approach that does purport to provide
estimates in such circumstances must raise concerns about its properties.

Consider a Weibull model for repeated measures with gamma random effects:
%
\begin{eqnarray}
\label{weibull1}
\quad f(\mathbf{y}_i|\bolds{\theta}_i)&=&
\prod_{j=1}^{n_i}\lambda\rho\theta_{ij}y_{ij}^{\rho-1}e^{\mathbf{x}
_{ij}'\mathbf{\xi}}
e^{-\lambda y_{ij}^{\rho}\theta_{ij}e^{\mathbf{x}_{ij}'\bolds{\xi
}}},
\\\label{weibull2}
f(\bolds{\theta}_i)&=&\prod_{j=1}^{n_i}\frac{1}{\beta_j^{\alpha
_j}\Gamma
(\alpha_j)}
\theta_{ij}^{\alpha_j-1}e^{-\theta_{ij}/\beta_j}.
\end{eqnarray}
Here, $i$ and $j$ are as in Section~\ref{unobservables}, $\theta
_{ij}$ are gamma random effects, $\mathbf{x}_{ij}$ are covariates,
$\bolds{\xi}$
regression parameters, $\rho$ the Weibull shape parameters, and
$\alpha_j$ and $\beta_j$ the parameters governing the gamma
distribution for the $j$th component.

Rather than the above two-parameter gamma density, it is customary in a
gamma frailty context\break (Ducha\-teau and Janssen, \citeyear{2007Duchateau}) to set $\alpha
_j\beta_j=1$, for reasons of identifiability.

In line with Bayarri's example, we use the less conventional constraint
$\alpha_j=1$ and $\beta_j=1/\delta_j$, leading to
%
\begin{equation}\label{alternativegamma}
f(\bolds{\theta}_i)=\prod_{j=1}^{n_i}\delta_je^{-\delta_j\theta_{ij}}
\end{equation}
and
implying that the gamma density is reduced to an exponential one.
Details can be found in Molenberghs et al.
(\citeyear{2009Molenberghset}).\vadjust{\goodbreak}

The moments take the following form:
%
\begin{eqnarray}\label{momentweibullexp}
 E(Y_{ij}^k)&=&
\frac{\delta_j^{k/\rho}k}{\lambda^k}\Gamma\biggl(1-\frac{k}{\rho
}\biggr)\Gamma\biggl(\frac{k}{\rho}\biggr)
\\
&&{}\times\exp
\biggl(-\frac{k}{\rho}\mathbf{x}_{ij}'\bolds{\xi}\biggr).\nonumber
\end{eqnarray}
Reducing the Weibull distribution to the exponential one, that is,
setting $\rho=1$, we further find
%
\begin{equation}\label{momentexpexp}
\ E(Y_{ij}^k)=
\frac{\delta_j^kk}{\lambda^k}\Gamma(1-k)\Gamma
(k)
\exp
(
-k\mathbf{x}_{ij}'\bolds{\xi}
).
\end{equation}
The cases corresponding to (\ref{momentweibullexp}) and, especially,
(\ref{momentexpexp}) will allow us to make our point about Lee and
Nelder's example. Generally, $\Gamma(\alpha-k/\rho)$ poses a problem
when $\alpha-k/\rho$ is a negative integer. For simplicity focusing
on a single outcome $Y$ for the case where $\alpha=1$ and $\beta
=1/\delta$, assembling the linear predictor in $\mu$, and writing
$\varphi=\lambda e^{\mu}$, we find
%
\begin{eqnarray}
\label{weibexp01}
f(y)&=&\frac{\varphi\rho y^{\rho-1}\delta}{(\delta+\varphi y^\rho
)^2},
\\\label{weibexp02}
E(Y^k)&=&\frac{k}{\rho}\biggl(\frac{\delta}{\varphi}\biggr)^{k\/
\rho}
\cdot\Gamma(1-k/\rho)\cdot\Gamma(k/\rho).
\end{eqnarray}
Note that (\ref{weibexp01}) provides a family of distributions,
special cases of the Weibull-gamma model that are termed
Weibull-exponential by Molenberghs et al. (\citeyear{2009Molenberghset}). Considering further
the exponential case with $\rho=1$, yields exponential-exponential
distributions, with
%
\begin{eqnarray}
\label{expexp01}
f(y)&=&\frac{\varphi\delta}{(\delta+\varphi y)^2},
\\\label{expexp02}
E(Y^k)&=&k\biggl(\frac{\delta}{\varphi}\biggr)^{k}
\cdot\Gamma(1-k)\cdot\Gamma(k).
\end{eqnarray}
Clearly, (\ref{expexp01}) defines a family of distributions without
finite moments similar to the Cauchy distribution because $\Gamma
(1-k)$ is undefined for $k=1,2,\ldots.$ When $\rho\ne1$ but is
fractional, some but not all moments exist whereas for irrational
values of $\rho$, all moments in (\ref{weibexp02}) are properly
defined. Finally, observe that in the general case, there are
combinations possible for $(\alpha,\rho,k)$ that would lead to
negative integers and hence undefined moments (\ref{momentweibullexp}).

In the light of the above developments, we are concerned that Lee and
Nelder provide us with point estimates for moments that are undefined.

\section{The Nature of Generalized Estimating Equations}\label{gee}

Lee and Nelder touch upon the use of generalized estimating equations
(GEE), as opposed to fully specified probability models. We agree that
a comparison between GEE to generalized linear mixed models (GLMM) or
hierarchical generalized linear models (HGLM) is like a comparison of
apples to oranges because GEE is an estimating method which can be
applied to random-effects models too (Zeger, Liang and Albert, \citeyear{1988Zeger})
and to HGLM for that matter. Therefore, again in agreement with Lee and
Nelder, we would like to reiterate that a proper basis of comparison is
between marginal and random-effects models. Thus, while part of the
literature is sloppy in making comparisons using sloppy
categorizations, it should not deflect from the real issues.
Nevertheless, we would like to reflect on whether GEE are indeed less
appealing because of their alleged lack of a probabilistic basis.

There are two important points in our view. First, a fully specified
probability model is not always essential when making inferences about
a particular aspect of the model, such as mean functions, as long as
the appropriate regularity conditions are satisfied. For example, when
estimating mean parameters, it is imperative that the mean exists and
is finite.

Second, as Molenberghs and Kenward (\citeyear{2008MolenberghsKe}) point\-ed out, the lower-order
moments that need to be formulated when setting up GEE, correspond to
at least one fully specified probabilistic model, even though it may
not be of the simple, elegant type, one has in mind {a priori},
such as the Bahadur (\citeyear{1961Bahadur}) or odds-ratio model (Molenberghs and
Lesaffre, \citeyear{1994Molenberghs}). Their work addresses the concern regarding whether
the model portions used in GEE can always be viewed as a partially
specified version of a model with full distributional assumptions, or
rather, whether such a {\em parent\/} simply does not exist. To this
end, they use the hybrid models (in the sense of being partially
marginally and partially conditionally specified) of Fitzmaurice and
Laird (\citeyear{1993Laird}) and Molenberghs and Ritter (\citeyear{1996Molenberghs}). The results by
Molenberghs and Kenward (\citeyear{2008MolenberghsKe}) are broadly valid. First, they are valid
for a wide class of semi-parametric models where specification is done
in terms of (parts of) the exponential-family formulation, including
binary, nominal, ordinal, and Poisson outcomes. Second, it is also
valid when the outcome vector combines outcomes of different types.
Third, using transformations, the result can be applied as well when
the semi-parametric specification is not directly in terms of the
exponential family, such as logistic regressions for binary data
coupled with pairwise correlation, as in classical generalized
estimating equations.

\section{Computational Approaches}\label{computation}

We are convinced that $h$-likelihood is a tremendously appealing and
important addition to the literature. Other computational principles
and techniques for maximizing a likelihood with unobservables exist as
well. Each one of them has its advantages and drawbacks. For example,
Taylor-series-based methods, such as PQL and MQL, and Laplace
approximations often lead to substantial bias. This is important to
realize because they have been in common use, nevertheless, not in the
least because they are implemented in standard statistical software
such as the SAS procedure GLIMMIX.\break The numerical-integration-based
methodology,\break implemented for example in the SAS procedure\break NLMIXED, is
frequently slow and/or extremely sensitive to starting values. See also
Molenberghs and Verbeke (\citeyear{2005Molenberghs}). Also Bayesian methods, sometimes
believed to be free of the issues arising in a likelihood or
frequentist context, also have their problems. For one, the sensitivity
arising from unobservables, as discussed in Section~\ref
{unobservables}, is equally present in this framework. It is clear to
us that none of the computational approaches will be able to claim
uniform superiority over all others.

Molenberghs et al. (\citeyear{2009Molenberghset}) provide an overview of computational methods,
including some less familiar ones. Their context is a hierarchical
model with both normally distributed and conjugate or other random
effects. Each of them deals in its own way with the lack of closed-form
expression for the marginal likelihood, even though Molenberghs et al.
(\citeyear{2009Molenberghset}) derive such closed forms for more settings such as general
Poisson, probit and Weibull models with random effects.

One approach, very specific to the setting of Molenberghs et al.
(\citeyear{2009Molenberghset}), is to integrate analytically over conjugate random effects and
then further numerically over the normally distributed random effects.

For the specific case of the marginalized probit model, the
computational challenge stems from the presence of a high-dimensional
multivariate normal integral in the marginal distribution. Zeger,
Liang and Albert (\citeyear{1988Zeger}) derived the marginal mean function, needed for
their application of generalized estimating equations as a fitting
algorithm for the marginalized probit model. It is one of the first
instances of the use of GEE to a nonmarginally specified model.

In the same spirit, pseudo-likelihood can be used (Aerts et al.,
\citeyear{2002Aerts};
Molenberghs and Verbeke, \citeyear{2005Molenberghs}). This is particularly useful when the
joint marginal distribution is available but cumbersome to manipulate
and evaluate, such as in the probit case. This is the idea followed by
Renard, Molenberghs, and Geys (\citeyear{2004Renard}) for a multilevel probit model with
random effects.

Schall (\citeyear{1991Schall}) proposed an efficient and general estimation algorithm,
based on Harville's (\citeyear{1974Harville}) modification of Henderson's (\citeyear{1984Henderson})
mixed-model equations. Hedeker and Gibbons (\citeyear{1994Hedeker}) and Gibbons and
Hedeker (\citeyear{1997Gibbons}) proposed numerical-integration based methods, thus
considering neither marginal moments (means, variances) nor
marginalized joint probabilities. Guilkey and Murphy (\citeyear{1993Guilkey}) provide a
useful early overview of estimation methods and then revert to Butler
and Moffit's (\citeyear{1982Butler}) Hermite-integration based method, supplemented with
Monte Carlo Markov chain ideas.
Also the EM algorithm can be used, in line with Booth et al. (\citeyear{2003Booth}) for
the Poisson case. The EM is a flexible framework within which random
effects can be considered the ``missing'' data over which expectations
are taken. Booth et al. (\citeyear{2003Booth}) also considered nonparametric maximum
likelihood, in the spirit of Aitken (\citeyear{1999Aitkin}) and Alf\`{o}\ and Aitkin (\citeyear{2000Alf}).

A suite of methods is available that employ transformation results,
essentially based on transforming the nonnormal random effects to
normal ones, or vice versa.
Liu and Yu (\citeyear{2008Liu}) propose a simple transformation of a nonnormal
random effect to a normal one, at density level, upon which the SAS
procedure NLMIXED or similar software can be used. Nelson et al. (\citeyear{2006Nelson})
advocate the transformation, $u_i=F_u^{-1}[(\Phi(a_i)]$ where $F_u$ is
the cumulative distribution function (CDF) of $u_i$, and $\Phi(\cdot
)$ is the standard normal CDF, as before. The method of Nelson et al.,
labeled {\em probability integral transformation\/} (P.I.T.), comes
down to generating normal variates and then inserting these in the
model only after transformation, ensuring that they are of the desired
nature. Lin and Lee (\citeyear{2008Lin}) present estimation methods for the specific
case of linear mixed models with skew-normal, rather than normal,
random effects.

Quite apart from the choice of estimation method, it is important to
realize that not all parameters may be simultaneously identifiable. For
example, the gamma-distribution parameters in the Poisson case, $\alpha
$ and $\beta$, such as in (\ref{weibull2}), are not simultaneously
identifiable when the linear-predictor part is also present because
there is aliasing with the intercept term. Therefore, one can set, for
example, $\beta$ equal to a constant, removing the identifiability
problem. It is then clear that $\alpha$, in the univariate case, or
the set of $\alpha_j$ in the repeated-measures case, describes the
additional overdispersion in addition to what stems from the normal
random effect(s). A similar phenomenon also plays in the binary case,
where both beta-distribution parameters are not simultaneously estimable.

\section{Concluding Remarks}

We end by sincerely thanking Professors Lee and Nelder for their
important, thought-provoking, and practically relevant work in this
rapidly evolving area. Their paper has allowed us to elaborate on a
number of statistical issues put forward in their paper, such as the
implications of formulating models with unobservables and generating
distributions with peculiar moment-properties. It further gave us the
opportunity to reflect on some conceptual aspects of generalized
estimating equations on one hand, and elaborate on computational
strategies for models of the type discussed here on the other.

\end{document}